\documentclass{article}
\usepackage{arxiv}

\usepackage[utf8]{inputenc} 
\usepackage[T1]{fontenc}    
\usepackage{hyperref}       
\usepackage{url}            
\usepackage{booktabs}       
\usepackage{amsfonts}       
\usepackage{nicefrac}       
\usepackage{microtype}      
\usepackage{lipsum}		
\usepackage{graphicx}

\usepackage{array}
\usepackage{amssymb, amsmath, amsthm}
\usepackage{graphicx}
\usepackage{lmodern,url}
\usepackage{makecell} 
\usepackage{cancel}
\usepackage{multirow}
\usepackage{microtype}
\usepackage{lscape}
\usepackage{xspace}
\usepackage{xcolor}
\usepackage{siunitx}
\usepackage{todonotes}
\usepackage{booktabs}
\usepackage[ruled,vlined]{algorithm2e}
\usepackage{optidef}
\usepackage{mathdots}
\usepackage{subcaption}
\usepackage{csquotes}
\usepackage{caption}
\usepackage{verbatim}
\usepackage{ulem}
\graphicspath{{Figures/}}

\usepackage{amsmath}
\usepackage{tikz}
\usetikzlibrary{fit}
\usetikzlibrary{positioning}
\tikzset{%
  highlight/.style={rectangle,rounded corners,fill=red!15,draw,fill opacity=0.25,thick,inner sep=0pt}
}

\definecolor{mygreen}{RGB}{160, 242,182}

\newcommand{\xunderbrace}[2][\vphantom{\dfrac{A}{A}}]{\underbrace{#1#2}}
\definecolor{mypurple}{RGB}{236, 223, 234}
\newcommand{\PrEP}{\ensuremath{P}\xspace}

\newcommand{\Iasti}{\ensuremath{I^{a}}\xspace}
\newcommand{\Issti}{\ensuremath{I^{s}}\xspace}
\newcommand{\Ssti}{\ensuremath{S}\xspace}
\newcommand{\Tsti}{\ensuremath{T}\xspace}
\newcommand{\Nobs}{\ensuremath{N^{\textrm{obs}}}\xspace}
\newcommand{\Nreal}{\ensuremath{N}\xspace}

\allowdisplaybreaks

\title{Testing Paradox May Explain Increased Observed Prevalence of Bacterial STIs among MSM on HIV P$\textnormal{r}$EP: A Modeling Study}

\usepackage{authblk}

\author[1,2$\dagger$]{Laura Müller}
\author[1,2$\dagger$]{Piklu Mallick}
\author[1,3$\dagger$]{Antonio B. Marín-Carballo}
\author[1,2$\dagger$]{Philipp Dönges}
\author[4,5]{Robyn J. N. Kettlitz}
\author[4]{Carolina J. Klett-Tammen}
\author[6,7,8]{Mirjam Kretzschmar}
\author[1,2]{Viola Priesemann}
\author[1,2*]{Seba Contreras}

\affil[1]{Max Planck Institute for Dynamics and Self-Organization, G\"ottingen, Germany.}
\affil[2]{Institute for the Dynamics of Complex Systems, University of G\"ottingen, G\"ottingen, Germany.}
\affil[3]{University of Granada, Granada, Spain.}
\affil[4]{Department of Epidemiology, Helmholtz Centre for Infection Research, Brunswick, Lower Saxony, Germany.}
\affil[5]{PhD Program “Epidemiology” Brunswick-Hanover, Germany}
\affil[6]{University Medical Center Utrecht, Utrecht University, Utrecht, The Netherlands.}
\affil[7]{Center for Complex Systems Studies (CCSS), Utrecht University, Utrecht, The Netherlands}
\affil[8]{Interdisciplinary Center for the Mathematical Modeling of Infectious Disease Dynamics (IMMIDD), University of Münster, Germany.}
\affil[ ]{* Corresponding Author: Seba Contreras (seba.contreras@ds.mpg.de)}
\affil[ ]{{$\dagger$} These authors contributed equally}

\date{}

\renewcommand{\headeright}{ }
\renewcommand{\undertitle}{ }

\begin{document}
\maketitle
\vspace*{-1cm}
\begin{abstract}
HIV pre-exposure Prophylaxis (PrEP) has become essential for global HIV control, but its implementation coincides with rising bacterial STI rates among men who have sex with men (MSM). While risk-compensation behavioral changes like reduced condom use are frequently reported, we examine whether intensified asymptomatic screening in PrEP programs creates surveillance artifacts that could be misinterpreted. We developed a compartmental model to represent the simultaneous spread of HIV and chlamydia (as an example of a curable STI), integrating three mechanisms: 1) risk-mediated self-protective behavior, 2) condom use reduction post-PrEP initiation, and 3) PrEP-related asymptomatic STI screening. Increasing PrEP uptake may help to reduce chlamydia prevalence, only if the PrEP-related screening is frequent enough. Otherwise, the effect of PrEP can be disadvantageous, as the drop in self-protective actions caused by larger PrEP uptake cannot be compensated for. Additionally, the change in testing behavior may lead to situations where the trend in the number of positive tests is not a reliable sign of the actual dynamics. We found a plausible mechanism to reconcile conflicting observational evidence on the effect of PrEP on STI rates, showing that simultaneous changes in testing and spreading rates may generate conflicting signals, i.e., that observed trends increase while true prevalence decreases. Asymptomatic screening, together with personalized infection treatment to minimize putative pressure to generate antibiotic resistance, is one of the key determinants of the positive side effects of PrEP in reducing STI incidence. 
\end{abstract}

 \clearpage
\section*{Introduction}

HIV pre-exposure prophylaxis (PrEP) has been transformative in reducing HIV infection risk for users \cite{anderson2012emtricitabine,mccormack2016pre}. Given its high efficacy and effectiveness at preventing HIV acquisition, high PrEP coverage is associated with substantial decreases in HIV incidence among risk groups, especially men who have sex with men (MSM) \cite{rozhnova2018elimination,murchu2022oral}. However, behavioral changes driven by risk compensation among MSM on PrEP make them more likely to engage in risky sexual behaviors, heightening their vulnerability to acquiring other sexually transmitted infections (STIs) deemed less severe (e.g., chlamydia, gonorrhea, trichomoniasis, and syphilis) \cite{hoornenborg2018change, holt2018community}. Although these are curable (in principle, see e.g., antibiotic-resistant strains of gonorrhea \cite{eyre2018gonorrhoea}), higher prevalence of such common STIs may drastically change the HIV infection risk landscape, especially in populations with low PrEP adherence \cite{galvin2004role,quaife2020risk}.

The shared transmission routes of many STIs further complicate prevention efforts due to the sporadic, private nature of these interactions, which can often be difficult to address from a public health perspective \cite{mogaka2023challenges}. As a result, efforts to control STI transmission heavily depend on individual behavioral changes and compliance \cite{kamb1998efficacy, costa2022hiv}. Asymptomatic screening for common STIs is part of the PrEP official recommendations and aims to mitigate the risk compensation among PrEP users \cite{spinner2019summary,jongen2023can,quaife2020risk}. Infectious individuals that get detected this way receive antibiotic treatment and recover, thereby reducing the force of infection \cite{contreras2021challenges}. However, the increasing number of detected STI infections as a consequence of the increased testing efforts may prompt associations between the use of PrEP and an increased incidence of STIs \cite{van2020diverging, jansen2020sti, traeger2019association, werner2018incidence}. This duality between \textit{true} and \textit{observed} trends (i.e., positive tests) challenges the interpretation of epidemiological data and thereby the assessment of PrEP as a public health intervention \cite{kim2023determining, ong2019global}.

In this study, we will examine two critical questions related to the effect of PrEP on STI prevalence among high-infection-risk groups of MSM in high-income settings (i.e., high/unrestricted PrEP uptake): (1) Can PrEP-related mandatory asymptomatic screening for common STIs effectively compensate for the drop in condom use among PrEP users? (2) When does screening lead to paradoxical dynamics, where the actual prevalence of STIs declines while the number of positive tests increases? By developing a mechanistic model, we aim to investigate the interactions between PrEP uptake, risk awareness, and screening frequency, specifically focusing on their combined impact on Chlamydia transmission among MSM. Our findings offer insights to improve the planning and evaluation of intervention strategies and help navigate the complexities of interpreting surveillance data.

\newpage

\section*{Methods}

We propose a deterministic compartmental model to represent the simultaneous spread of HIV and Chlamydia in a high-infection-risk group of MSM, including the effects of PrEP, HIV infection risk perception, and asymptomatic screening for STIs. The dynamics in this multi-pathogen system are governed by two key variables: the overall PrEP uptake $P$ and the HIV-infection-risk awareness $H$ among the high-infection-risk population. In our framework, individuals on PrEP fully comply with the screening requirements and administration recommendations. High infection risk is defined as a high number of new sexual partners and overall low use of condoms, which translates to high base spreading rates for STIs. 
 
We hypothesize that individuals who are not on PrEP adapt their behavior according to their perceived HIV infection risk \cite{wagner2025societal,donges2022interplay}. This adaptation implies that these individuals would adopt mitigation measures $m\left(H\right)$ to reduce the risk of infection upon contact and seek asymptomatic STI screening at a rate $\lambda_H\left(H\right)$, both monotonically increasing with $H$. PrEP users, on the other hand, behave differently. We assume that they change their sexual behavior by not engaging in additional mitigation measures and not showing a marked response to HIV infection risk, as they are protected. However, public health plans require them to test for HIV and other STIs at a rate $\lambda_P$, which varies from country to country \cite{prins2023statutory,spinner2019summary}. That way, PrEP uptake acts in a positive feedback loop with the effective transmission rate of curable STIs and the total testing rate of STI infections.

Our model incorporates the possibility that Chlamydia can manifest either as notably symptomatic or asymptomatic. We assume that asymptomatic individuals are detected through testing only if they undergo screening as a response to their risk awareness ($\lambda_H$) or as part of routine PrEP-related screening ($\lambda_P$). On the other hand, symptomatic individuals additionally seek testing due to symptom onset at a rate $\lambda_0$, which is defined as the inverse of the median incubation period ($1/l_{STI}$). In either case, once detected, positive individuals receive treatment (antibiotics and counseling) so that they are effectively removed from the pool of infections and recover. We also assume that there is a short period of immunity after treatment, either due to immune response or behavioral changes. 

To incorporate demographic dynamics into the model, we introduce a death rate $\mu$ and a recruitment rate $\Phi$. The recruitment term represents the entry of new individuals into the high-infection-risk MSM population and serves a similar role to a birth rate in traditional models. The condition $\mu = \Phi$ yields a demographic equilibrium stabilizing the population size.

Lastly, although we employ parameters based on Chlamydia, our analysis seeks to shed light on the transmission of any of the four curable STIs (chlamydia, gonorrhea, trichomoniasis, and syphilis), which, for modeling purposes, differ only in parameter values. Hereafter, sub-indexes and references to STI in our model refer to chlamydia.

\subsection*{Model equations}

\begin{figure}[!h]
    \centering    \includegraphics[width=100mm]{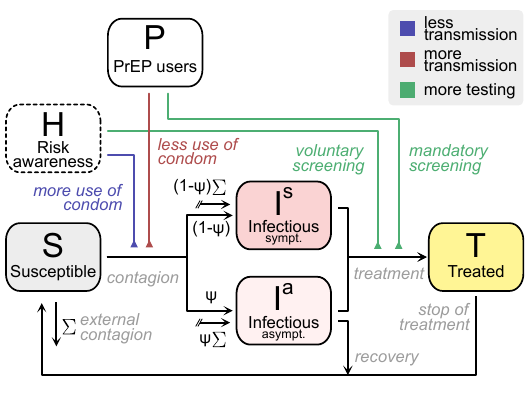}
    \caption{%
        \textbf{Minimal susceptible-infectious-treated-susceptible model for chlamydia transmission among high-risk MSM, influenced by HIV risk perception and PrEP uptake.} This model distinguishes between symptomatic and asymptomatic infections and incorporates three feedback mechanisms: i) increased risk awareness leads to higher condom use (blue, reducing transmission rates), ii) lower condom use among PrEP users (red, increasing transmission rates), and iii) enhanced asymptomatic screening due to both risk perception and PrEP uptake (green, increasing testing rates). Our model also incorporates an additive influx of externally-acquired infections $\Sigma$, distributed between the \Iasti and \Issti proportional to the share of cases that are naturally asymptomatic $\psi$ (or its complement).
        }
    \label{fig:minimal_model}
\end{figure}

Given that the spreading dynamics of HIV and chlamydia have different characteristic times (e.g., as per their generation intervals and spreading rates), we resort to the method of timescale separation to simplify our analysis: When the focus is on the dynamics of chlamydia, HIV incidence and awareness do not vary substantially. As a result, we treat the fraction of the population that is on PrEP $P$ and HIV prevalence, thus the population’s infection risk awareness $H$, as constant inputs when simulating the STI dynamics. A flow diagram of the model is shown in Fig.~\ref{fig:minimal_model}. 

\begin{align}
\frac{d \Ssti}{dt}    & =  -\xunderbrace{\Lambda\Ssti}_{\text{contagion}}+\xunderbrace{ \gamma \Iasti + \tilde{\gamma} \Tsti}_{\text{immunity loss}}  + \xunderbrace{\Phi }_{\text{recruitment rate}} - \xunderbrace{\mu \Ssti}_{\text{deaths}} - \xunderbrace{\Sigma}_{\text{total influx}}  & : \quad& \text{susceptible,}\label{eq:dSsti_simple}\\
\frac{d \Iasti}{dt}    & =  \xunderbrace{\psi \cdot \Lambda\Ssti}_{\text{asymp. contagion}} - \xunderbrace{\gamma \Iasti}_{\text{recovery}} - \xunderbrace{\lambda_a(P,H)\Iasti}_{\text{testing and treatment}} - \xunderbrace{\mu \Iasti}_{\text{deaths}} + \xunderbrace{\psi\Sigma}_{\text{asymp. influx}}   & : \quad& \text{asymptomatic infection,} \label{eq:dIasti_simple}\\
\frac{d \Issti}{dt}    & =  \xunderbrace{(1-\psi) \cdot \Lambda\Ssti}_{\text{symptomatic contagion}} - \xunderbrace{\lambda_s(P,H)\Issti}_{\text{testing and treatment}} - \xunderbrace{\mu \Issti}_{\text{deaths}} + \xunderbrace{(1-\psi)\Sigma}_{\text{symptomatic influx}}   & : \quad& \text{symptomatic infection,}\label{eq:dIssti_simple}\\
\frac{d \Tsti}{dt}    & =  \xunderbrace{\lambda_a(P,H) \Iasti}_{\text{testing and treatment}} + \xunderbrace{\lambda_s(P,H)\Issti}_{\text{testing and treatment}} - \xunderbrace{\tilde{\gamma} \Tsti}_{\text{recovery}} - \xunderbrace{\mu \Tsti}_{\text{deaths}}  & : \quad& \text{treated/recovered,}  \label{eq:dTsti_simple}
\end{align}

\begin{align}
        \Lambda & = \beta_{0}^{\textrm{STI}}\left( (1 - m(H))(1 - P) + P \right )(\Iasti+\Issti) & : \quad& \text{force of infection,}\\
    \lambda_{a}(P,H) & = \xunderbrace{\lambda_H(H) \cdot (1-\PrEP)}_{\text{risk-related testing | self-reporting}} + \xunderbrace{\lambda_P \cdot \PrEP}_{\text{PrEP-related testing rate}} & : \quad& \text{asymptomatic testing rate,} \label{eq:feedback_3}\\
\lambda_{s}(P,H) & = \xunderbrace{\lambda_0}_{\text{self-reporting rate}} + \xunderbrace{\lambda_{a}(P,H),}_{\text{testing rate for asymptomatic}}& : \quad& \text{symptomatic testing rate,} \label{eq:feedback_4}\\
\lambda_H(H) & =  k_\text{or}  \cdot \beta^{\text{HIV}}_0 (1-m(H))  H& : \quad& \text{risk-related testing rate,} \label{eq:feedback_5}\\
m(H) & =  m_{\textrm{min}} + (m_{\textrm{max}}-m_{\textrm{min}})\left(1-\exp\left(\frac{-H}{H_{\textrm{max}}}\right)\right)& : \quad& \text{mitigation function.} \label{eq:mitigation}
\end{align}

For the main text of this article, we focus our analysis on a single high-infection-risk group of MSM in high-income settings, where the effective PrEP uptake is limited only by willingness and compliance. This allows obtaining analytical solutions for the steady state of the system in closed form and gaining insights into the fundamental mechanisms behind the observed dynamics. Readers are referred to the supplementary material for an extended analysis using a complex, risk-stratified model for STI spread.

\subsection*{Central epidemiological variables that can be observed}

Given the modeling nature of our study, we can determine the exact fraction of the population infected (and infectious) with an STI—i.e., the total prevalence—at any point in time. In reality, however, prevalence can only be estimated through testing. To account for this, we introduce \Nreal and \Nobs. The \textit{true} prevalence, \Nreal, represents the total number of active infections, whereas the \textit{observed} cases, \Nobs—i.e., the number of positive test results—depend on the testing rates $\lambda_s$ and $\lambda_a$, as well as on \Nreal.

\begin{align}
\Nreal & =\Issti+\Iasti & : \quad & \text{Total prevalence (active cases),} \label{eq:Nreal}\\
\Nobs & = \lambda_s\Issti + \lambda_a\Iasti & : \quad & \text{Daily positive tests.} \label{eq:Nobs} 
\end{align}

\begin{table*}[ht!]\caption{\textbf{Model variables.} All variables represent fractions of the population unless stated otherwise.}
\label{tab:variables}
\centering
\begin{tabular}{lp{10cm}}\toprule
\makecell[l]{Variable} & Definition  \\\midrule
$\Ssti$ & Fraction of the population susceptible to curable STIs  \\
$\Iasti$ & Fraction of the asymptomatic population infectious with curable STIs  \\
$\Issti$ & Fraction of the symptomatic population infectious with curable STIs  \\
$\Tsti$ & Fraction of the population treated against curable STIs  \\
$H$ & Fraction of the population that is risk aware  \\
$P$ & Fraction of the population that is on HIV PrEP  \\
$N$ & STI Prevalence (fraction of the population currently infectious), Eq. \ref{eq:Nreal}  \\
\Nobs & Observed STI cases (daily positive tests), Eq. \ref{eq:Nobs}  \\
\bottomrule
\end{tabular}%
\end{table*}

\begin{table*}[ht!]\caption{\textbf{Model parameters.}}
\label{tab:parameters}
\begin{tabular}{lp{8cm}lll}\toprule
Parameter & Definition   & Value & Units & Source \\\midrule
$\beta_{0}^{\textrm{HIV}}$ & Base spreading rate of HIV within risk group & $0.6341$ & \SI{}{yr^{-1}}& \cite{rozhnova2018elimination}\\
$\beta_{0}^{\textrm{STI}}$ & Base spreading rate of Chlamydia within risk group &  $\{0.008,0.0112\}$& \SI{}{day^{-1}} & \cite{montes2022evaluating}\\
$\gamma$ & Chlamydia recovery rate (natural) & $1/1.32$ & \SI{}{yr^{-1}} & \cite{qu2021effect}\\
$\tilde{\gamma}$ & Chlamydia treatment-mediated recovery rate & 1/7 & \SI{}{day^{-1}} & \cite{qu2021effect}\\
$\Sigma$ & Total influx people infected with STI & $0.01$ & \SI{}{yr^{-1}} & Assumed\\
$\psi$ & Fraction of asymptomatic Chlamydia infection & $0.85$ & - & \cite{lamontagne2003chlamydia,korenromp2002proportion} \\ 
$\Phi$ & Recruitment rate to sexually active population & 1/45 & \SI{}{yr^{-1}} &\cite{rozhnova2018elimination} \\
$\mu$ & Exit rate from sexually active population & 1/45 & \SI{}{yr^{-1}} &\cite{rozhnova2018elimination} \\
$l_{\textrm{STI}}$ & Incubation period for Chlamydia infection in men & 14 & \SI{}{day} & \cite{korenromp2002proportion} \\
$\lambda_0$ & Self-reporting rate for symptomatic STI infection & $1/l_{\textrm{STI}}$ & \SI{}{day^{-1}} & Calculated \\
$\lambda_P$ & PrEP-related screening rate for STI & [0,4] & \SI{}{yr^{-1}} & \cite{world2021consolidated,hevey2018prep}, scanned \\
$\lambda_a(P,H)$ & Testing rate (asymptomatic) for STI & eq. \ref{eq:feedback_3}& \SI{}{day^{-1}} & Calculated\\
$\lambda_s(P,H)$ & Testing rate (symptomatic) for STI & eq. \ref{eq:feedback_4}& \SI{}{day^{-1}} & Calculated\\
$\lambda_H (H)$ & Risk-related testing rate for STI or self-reporting & eq. \ref{eq:feedback_5}& \SI{}{day^{-1}} & Calculated\\
$k_\text{or}$ & Odds ratio of perceiving risk if one is at risk & 50 & - & \cite{kesler2016actual}\\
$m $ & Mitigation, self-regulation of contagious contacts & eq.~\ref{eq:mitigation} & - & Calculated\\
$m_{\textrm{min}}$ & Minimum mitigation & 0  & - & Assumed \\
$m_{\textrm{max}}$ & Maximum mitigation & 1  & - & Assumed \\
$H_{\textrm{max}}$ & Characteristic reaction awareness & $0.2$  & \SI{}{-} & Assumed\\
\bottomrule
\end{tabular}%
\end{table*}

\clearpage
\section*{Results}
\subsection*{Increasing PrEP uptake can reduce the prevalence of chlamydia}

Increasing the uptake of PrEP, despite its association with reduced condom use, can effectively lower the prevalence of chlamydia and other curable STIs among high-infection-risk groups. This seemingly counterintuitive outcome is explained by the necessity for individuals receiving PrEP to adhere to a regular screening schedule, which includes asymptomatic STI screening. Frequent screening enables the early detection and treatment of infectious but asymptomatic individuals, thereby reducing the time they contribute to disease transmission. Testing, as an active intervention, decreases the effective force of infection and lowers the overall reproduction number, potentially bringing it below one.

\begin{figure}[!h]
\hspace*{0cm}
    \centering
    \includegraphics[width=165mm]{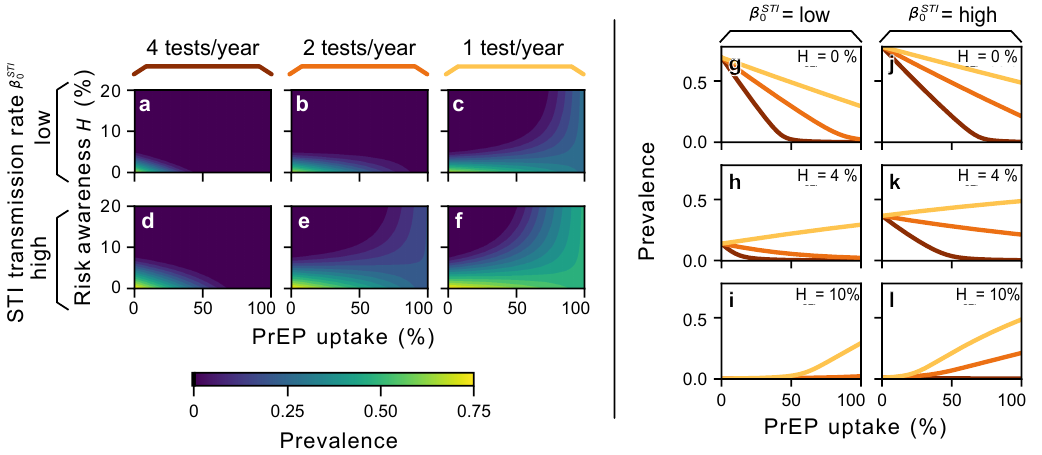}
    \caption{%
        \textbf{Increasing PrEP use among high-infection-risk groups can mitigate the spread of chlamydia.}  \textbf{a--f} We analyze the expected prevalence of chlamydia infections under two scenarios of transmission rates (low and high; parameters in Tab.~\ref{tab:parameters}). The frequency of PrEP-related screening determines the effects of PrEP uptake, particularly at low levels of risk awareness in the population ($H$). For instance, when PrEP users are required to undergo STI screening every three months (\textbf{a}, \textbf{d}), higher PrEP uptake effectively reduces the expected prevalence. This benefit diminishes as screening frequency decreases. \textbf{g--l} The relationship between risk awareness and PrEP uptake is complex; increasing PrEP uptake results in a reduction of the fraction of the population that is sensitive to $H$ and thereby reduces the relative impact of their self-protective actions. At low risk awareness, PrEP screening dominates, and the overall effect of PrEP is beneficial (\textbf{g}), but this effect can reverse when awareness is higher (\textbf{h}, \textbf{i}). This effect is more pronounced when the STI transmission rate is high (\textbf{j--l}).
        }
    \label{fig:results_1}
\end{figure}

We analyze two scenarios regarding the transmission rate of chlamydia alongside three different (PrEP-related) screening frequencies within high-infection-risk groups of MSM (Fig.~\ref{fig:results_1}). At high screening frequencies (i.e., 4 tests per year), increased PrEP uptake correlates with a reduction in chlamydia prevalence. However, with decreasing screening frequency, a larger increase in PrEP uptake is necessary to prevent an increase in prevalence. Furthermore, if the screening frequency falls below a critical threshold, the net effect of PrEP reverses to be detrimental --- asymptomatic screening is not sufficient to counteract the increased spreading rate caused by the drop in condom use among PrEP users (Fig.~\ref{fig:results_1}a--f, and Supplementary Fig.~\ref{fig:min_test_frequ}a--f). 

The relationship between risk awareness and PrEP uptake is complex; increasing PrEP uptake results in a reduction of the fraction of the population that is sensitive to $H$ and thereby reduces the relative impact of their self-protective actions. On the one hand, at very low risk awareness (i.e., when $H\approx 0$ ), the overall asymptomatic STI screening is dominated by the PrEP-related testing, even reaching the elimination threshold when a critical threshold of PrEP uptake is reached. However, this critical threshold is higher when screening occurs less frequently, which could render it unachievable if screening rates fall below a critical point (cf. to Fig.~\ref{fig:results_1}g). On the other hand, when $ H > 0\% $, the effect of the risk-perception-driven testing $\lambda_H$ may dominate the overall testing rates, thus requiring more frequent PrEP-related screening $\lambda_P$ to compensate in situations where PrEP use is widespread. In other words, larger PrEP uptake leads to higher chlamydia prevalence in situations where PrEP-related screening is low (see Fig. \ref{fig:results_1}h,i, and Supplementary Fig. \ref{fig:min_test_frequ}). This highlights the need for frequent STI screening for PrEP users, together with targeted information campaigns. 

Importantly, even for a lower fraction of asymptomatic infections ($\psi = 0.7$\cite{williams2023frequent}), the dynamics remain consistent: prevalence increases with PrEP uptake when PrEP-related testing is low and decreases when testing is frequent (Supplementary Fig.~\ref{fig:figure_1_psi_0.7}). The overall prevalence, however, decreases. This is because individuals with symptoms will voluntarily seek treatment and are thus removed from the infection pool. Asymptomatic screening, therefore, plays a smaller role, making the system less reliant on frequent asymptomatic testing to interrupt transmission.

\clearpage
\subsection*{Testing paradox: observing more but having less}

\begin{figure}[!h]
\hspace*{0cm}
    \centering
    \includegraphics[width=165mm]{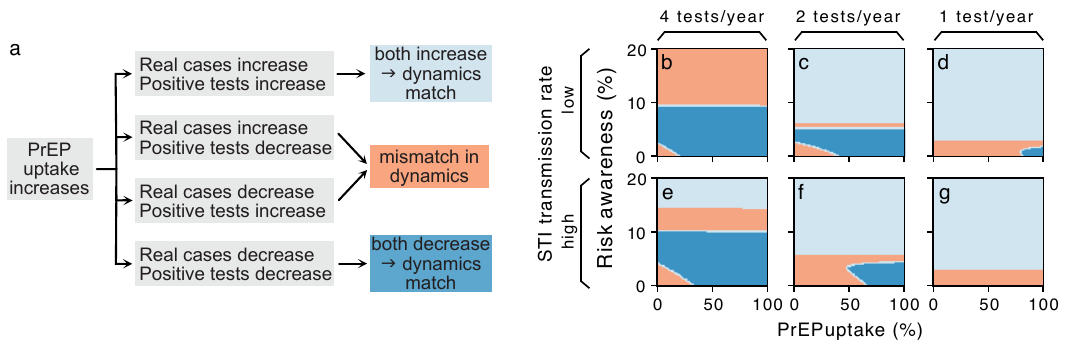}
        \caption{\textbf{Mismatch in the dynamics of observed and real cases.} \textbf{a} An increase in PrEP uptake will change the prevalence \Nreal and observed cases \Nobs. Both can either increase or decrease, resulting in four different combinations that can be assigned into three categories: (light blue) both \Nreal and \Nobs will increase as a consequence of increased PrEP uptake in the population, (dark blue) both \Nreal and \Nobs will decrease as a consequence of increased PrEP uptake in the population, (red) one of the two increases, while the other decreases. This last scenario leads to a mismatch in dynamics. This means that the observed dynamics (number of positive tests) do not represent the true trends of STI prevalence and thus give a false picture of what is happening. \textbf{b-g} Computing the product of the derivatives of \Nreal and \Nobs with respect to PrEP uptake ($\frac{\partial\Nreal}{\partial P}\cdot\frac{\partial\Nobs}{\partial P}$) unveils whether the change in observed cases matches the change in real cases. We found several regions with mismatches (red) where real cases decrease while observed cases rise. (Parameters in Tab.~\ref{tab:parameters})}
\label{fig:results_2}
\end{figure}

When analyzing how PrEP uptake and risk awareness in the population affect chlamydia spread, we must distinguish between true prevalence (\Nreal; Eq.~\eqref{eq:Nreal}) and observed cases (number of positive tests, \Nobs; Eq.~\eqref{eq:Nobs}). Typically, we assume that while \Nobs underestimates \Nreal, it is a reliable signal of its trends. However, while more frequent screening increases the number of cases uncovered, it also reduces the true prevalence, as these uncovered individuals are treated and do not contribute to spreading the disease further. This creates a paradoxical result; we may see more cases while the true prevalence declines. To understand when and why this paradox arises, we analyzed the rates of change in \Nreal and \Nobs with respect to increasing PrEP uptake (Fig.~\ref{fig:results_2}), i.e., how much do they change when increasing a unit of PrEP uptake. These rates are mathematically quantified by their partial derivatives ($\frac{\partial\Nreal}{\partial P}$, $\frac{\partial\Nobs}{\partial P}$). For a wide range of parameters, these derivatives have matching signs. Therefore, \Nobs is a reliable signal for the direction of changes in \Nreal (i.e., both increase or decrease together, respectively light and dark blue in Fig.~\ref{fig:results_2}). Regions where the product of the partial derivatives of \Nreal and \Nobs is negative (red) are where the paradox arises: one quantity increases (positive derivative) and the other decreases (negative derivative).

The contribution of PrEP to the overall testing rate goes beyond the mandatory screening: as a consequence of risk compensation among PrEP users, PrEP uptake shapes the testing behavior in the whole population. When increasing PrEP uptake, the fraction of the population that is not on PrEP (and thus reacts to the perceived HIV infection risk) decreases. This can lower the overall STI testing rate, especially when PrEP-related testing rates $\lambda_P$ are low. 
When the overall testing increases, previously undiscovered infections become uncovered, and people receive treatment.
If there are more susceptible than infectious individuals (i.e., $S>I$), the force of infection decreases, thereby reducing \Nreal. For the dynamics of \Nobs, two factors play a role: If \Nreal stays constant, increased testing leads to higher \Nobs simply because we test more. If the testing rate stays constant, on the other hand, an increase or decrease in \Nreal will lead to an increase or decrease in \Nobs, respectively, because the positivity rate changes. 
When both the positivity rate and the overall number of tests change simultaneously, the change in \Nobs depends on which of the two has the bigger effect.

If risk awareness and PrEP uptake both are low, increasing PrEP uptake will increase the overall testing rate, causing \Nreal to decrease and \Nobs to increase (lower left corners in Fig.~\ref{fig:results_2}b--f). 
On the other hand, increasing PrEP uptake at low risk awareness and high PrEP uptake will most likely increase the overall testing rate so that \Nreal decreases fast enough to cause a decline in \Nobs too. This effect is more pronounced when the PrEP-related screening rate is high. 
However, for medium risk awareness, risk-related and PrEP-related testing frequencies are similar, thus the overall testing increases only by a small amount as PrEP uptake increases. Consequently, \Nreal, and therefore the positivity rate, also only decreases by a small amount. The overall increase in testing, therefore, has a larger effect and leads to an increase in \Nobs. With higher PrEP-related testing frequency, this region starts at higher risk awareness levels and spans larger ranges of risk awareness (Fig.~\ref{fig:results_2}b,e). For lower testing rates, on the other hand, risk-related testing is higher than PrEP-related testing even at relatively low risk awareness levels, so that increasing PrEP uptake means less total testing, leading to an increase in both real and observed cases (Fig.~\ref{fig:results_2}c,d,f,g). Even though we do not observe it here, it could also happen that \Nreal increases while \Nobs decreases. This would be the case if testing decreases at fast levels, fueling new infections while at the same time observing fewer simply because we test less. Increasing PrEP uptake in the population can therefore have differing effects on \Nreal and \Nobs depending on multiple factors, including risk awareness and PrEP-related testing frequency.

Note that all analyses in this section, which focused on changing PrEP uptake in the population, can also be done with respect to changing risk awareness. This analysis, alongside robustness tests, can be found in the Supplementary section \ref{SI_sec:Nobs_Nreal_riskawareness}.

\clearpage
\section*{Discussion}

Our model suggests that increasing PrEP uptake can help to reduce the prevalence of chlamydia (as an example of a curable STI) in high-infection-risk groups of MSM. The mechanism behind this reduction is the asymptomatic STI screening: When an infection is detected, treatment is provided, so that individuals are cured or adapt their behavior following the detection of the pathogen, thereby reducing the effective force of infection and overall spreading rate. However, without regular and frequent screening, testing cannot counteract the risk compensation among PrEP users, and the secondary benefits of PrEP in controlling chlamydia prevalence may be reversed. Thus, regular and frequent screening should remain a fundamental part of PrEP guidelines. These findings are aligned with observational data \cite{streeck2022hiv,schmidt2023low} and other modeling studies \cite{jenness2017incidence}.

However, there is some debate in the field related to the benefits of frequent asymptomatic screening for some STIs~\cite{williams2023frequent}, especially for chlamydia and gonorrhea, arguing that there is no clear link between screening and reduced rates for these STIs among MSM \cite{tsoumanis2018screening}, and it may cause an increased use of broad-spectrum antibiotics \cite{vanbaelen2021screening}. First, observational data show conflicting evidence on whether more screening led to permanent reductions in the incidence of chlamydia and gonorrhea \cite{tsoumanis2018screening,vanbaelen2024effect}. We discovered a plausible mechanism to reconcile these results and demonstrated that, across a broad range of parameters, testing artifacts account for the observed rise in STI prevalence, even as the true prevalence declines. Second, although more frequent screening indeed leads to overexposure to antibiotics~\cite{vanbaelen2024effect}, we stress that this is not a consequence of asymptomatic screening but rather a lack of personalization in treating such infections: alternative and personalized antimicrobial regimens limiting the use of azithromycin should also be considered \cite{horner20162016}. Furthermore, the behavioral component of knowing one's health status is independent of the availability of a treatment: once one is aware of being infectious, reacting to protect others bears no or only negligible costs to the person \cite{schnyder2025understanding,kettlitz2023association}. 

Our model includes several simplifying assumptions, made to facilitate the interpretation of the dynamics and isolation of key mechanisms. First, we assume that HIV prevalence remains constant over time, justified by a separation of timescales: HIV spreads much more slowly than curable STIs like chlamydia. Testing behavior is modeled as being influenced by risk awareness, which we assume to be proportional to HIV prevalence. While individuals on PrEP are assumed to be less likely to test voluntarily (due to lower perceived risk), they are required to test a fixed number of times per year per public health guidelines~\cite{spinner2019summary}. We assume no functional or causal relationship between PrEP uptake and HIV prevalence. Instead, we consider multiple combinations of these two variables (PrEP uptake from $0$ to $100\%$ and risk awareness up to $20\%$), treating them as independent for the purpose of our analysis. To ensure our findings also hold for more realistic modeling choices, we also explored a more complex version of the model that stratifies the population into four validated infection-risk groups (Supplementary Materials, Section~\ref{SI_sec:extended_model}). In this extended framework, individuals differ in their levels of risk awareness, infection risk, and contact patterns between groups---features that more closely reflect real-world heterogeneity. Despite the substantial increase in complexity, results remain qualitatively consistent (cf. Fig.~\ref{SI_fig: figure1_scenarios123} to mirror Fig.~\ref{fig:results_1} and Fig.~\ref{SI_fig: figure2_scenarios123} to mirror Fig.~\ref{fig:results_2}. This suggests that the core insights derived from the minimal model are robust and general.

To sum up, the effectiveness of interventions like PrEP needs to be carefully assessed to rule out testing artifacts, as they affect both the dynamics and screening rates of STIs. We showed a plausible mechanism to reconcile conflicting observational evidence on the effects of PrEP on the prevalence of STIs, together with analyzing the sensitivity of such effects with regard to different levels of risk awareness, PrEP-related testing frequencies, and transmission rates. Wherever resources are available, asymptomatic screening together with personalized treatment of infections remains an attractive way to slow down the spread of STIs, including HIV~\cite{jenness2017incidence,jones2022effect}, especially in high-infection-risk groups. Individuals in these groups may act as bridges to other populations and settings where PrEP uptake is not widespread. Ultimately, increasing the prevalence of chlamydia and other curable STIs in these unprotected groups increases their susceptibility to acquiring HIV. 

\section*{Author Contributions}

Conceptualization: SC\\
Data curation: AM, LM, PD, PM\\
Formal analysis: AM, LM, PD, PM\\
Funding acquisition: SC, VP\\
Investigation: AM, CK-T, LM, MK, PM, RK\\
Methodology: AM, LM, PD, PM, SC\\
Project administration: SC, VP\\
Resources: VP\\
Software: AM, LM, PD, PM\\
Supervision: SC, VP\\	
Validation: all\\
Visualization: AM, LM, PM, SC \\	
Writing - Original Draft: AM, LM, PM, SC\\	
Writing - Review \& Editing: all \\

\section*{Code availability}
All code to reproduce the analysis and figures shown in the manuscript as well as in the Supplementary Material is available online on Github \href{https://github.com/Priesemann-Group/testing-paradox-HIV-STI.git}{https://github.com/Priesemann-Group/testing-paradox-HIV-STI}. 

\section*{Acknowledgments} 

We thank the Priesemann group, especially Dr Fabio Sartori, for the exciting discussions and valuable comments they provided.
Authors with affiliation "1"  received support from the Max-Planck Society. VP, SC, PD, CK-T were funded by the German Federal Ministry for Education and Research for the RESPINOW project (031L0298), and VP, SC, LM, PM by the infoXpand project (031L0300A). VP was supported by the Deutsche Forschungsgemeinschaft (DFG, German Research Foundation) under Germany’s Excellence Strategy - EXC 2067/1-390729940  (MBExC). ChatGPT and Grammarly AI were used for grammar checks in the main text, and GitHub Copilot served as a coding assistant. The authors assume full responsibility for the final content of the article.

\clearpage

\newpage
\renewcommand{\thefigure}{S\arabic{figure}}
\renewcommand{\figurename}{Supplementary~Figure}
\setcounter{figure}{0}
\renewcommand{\thetable}{S\arabic{table}}
\renewcommand{\tablename}{Supplementary~Table}

\setcounter{table}{0}
\renewcommand{\thesection}{S\arabic{section}}
\setcounter{section}{0}
\setcounter{page}{1}
\section{Supplementary Material}

\subsection{Fixed points of the simple HIV-STI model}

In our model, the population is divided into compartments corresponding to different health and disease states: susceptible ($\Ssti$), asymptomatic infectious ($\Iasti$), symptomatic infectious ($\Issti$), and on treatment ($T$). Each compartment represents the fraction of the total population in that category. We assume the total population remains constant over time and normalize it to 1:
\begin{equation}
    \Ssti + \Iasti + \Issti + \Tsti = 1.
\end{equation}

This condition, known as demographic equilibrium, implies that the sum of all derivatives is zero. This allows us to obtain an expression for the recruitment rate $\Phi$:
\begin{equation}                            
\Phi=\mu\left(\Ssti + \Iasti + \Issti + \Tsti\right)=\mu.
\label{eq:demographic_equilibrium}
\end{equation}

Additionally, it allows reducing the system to three differential equations, given that $\Tsti$ can be represented as an algebraic function of the other variables:
\begin{equation} 
\Tsti = 1 - \left(\Ssti + \Iasti + \Issti\right).
\end{equation}

We determine the existence of fixed points ($S^*$, $I_a^*$, $I_s^*$) by imposing the stability condition $\frac{d}{dt}\left(S^*, I_a^*, I_s^*\right)=(0,0,0)$. Substituting the conditions above, equations~\eqref{eq:dSsti_simple}--\eqref{eq:dIssti_simple} can be rewritten as follows:
\begin{align}
0 &= -C_1 \left(I_a^*+I_s^*\right) S^* + \gamma I_a^* + \tilde{\gamma} \left(1 - (S^* + I_a^* + I_s^*)\right) + \mu - \mu S^* - \Sigma \, ,\label{fpeq:1} \\
0 &= \psi C_1\left(I_a^*+I_s^*\right) S^* - (\gamma + \lambda_a) I_a^* - \mu I_a^* + \psi \Sigma\, , \label{fpeq:2} \\
0 &= (1 - \psi) C_1\left(I_a^*+I_s^*\right) S^* - (\lambda_s + \mu) I_s^* + (1 - \psi) \Sigma \label{fpeq:3}\, ,
\end{align}

where $C_1 = \beta_0^{\mathrm{STI}}\left(1-m(H)\left(1-P\right)\right)$. 

We calculate $ S^* $ from equation \eqref{fpeq:3}:
\begin{equation}
S^* = \frac{ (\lambda_s + \mu) I_s^* - (1 - \psi) \Sigma }{ (1 - \psi) C_1 (I_a^* + I_s^*) }\, .\label{fpeq:8}
\end{equation}

The relationship between $ I_s^* $ and $ I_a^* $ follows from equations \eqref{fpeq:2} and \eqref{fpeq:8}:
\begin{equation}
I_s^* = \frac{ (\gamma + \lambda_a + \mu) (1 - \psi) }{ \psi (\lambda_s + \mu) } I_a^*  \overset{\underset{\mathrm{def}}{}}{=} C_2  I_a^* \, .\label{fpeq:13}
\end{equation}

Using equations \eqref{fpeq:1}, \eqref{fpeq:8}, and \eqref{fpeq:13}, we obtain a quadratic equation for $ I_a^* $
\begin{equation}
    a(I_a^*)^2 + bI_a^* + c = 0\, ,\label{eq:quadratic_for_Ia}
\end{equation}

where
\begin{align}
a &= C_1 (1 + C_2) \left[ - (\lambda_s + \mu) C_2 + (\gamma - \tilde{\gamma} (1 + C_2))(1 - \psi) \right]\, , \label{fpeq:21} \\
b &= (\tilde{\gamma} + \mu) \left[C_1 (1 - \psi) (1 + C_2) - (\lambda_s + \mu) C_2 \right] \, ,\label{fpeq:22} \\
c &= (\tilde{\gamma} + \mu)(1 - \psi) \Sigma \, .\label{fpeq:23} \\
\end{align}

From this follows:

\begin{equation}
    I_a^* = \frac{ -b \pm \sqrt{b^2 - 4ac} }{ 2a }\, . \label{fpeq:24}
\end{equation}

Given that $C_1,\,C_2 > 0$, and assuming that $\tilde{\gamma}\left(1+C_2\right) > \gamma$ (which in the parameter regime we explore is true even for the case $C_2 = 0$), we can conclude that:
\begin{itemize}
    \item In Eq.~\ref{eq:quadratic_for_Ia}, $a<0$ and $c>0$.
    \item Both solutions are real, given that $b^2-4ac>0$ always.
    \item There is only one positive solution for $I_a^*$ in Eq.~\ref{fpeq:24}, given that the product of the roots of Eq.~\ref{eq:quadratic_for_Ia} fulfills $x_1x_2 = c/a <0$.
\end{itemize}
We therefore showed the existence of one positive real-valued fixed point.

\subsection{Minimum PrEP-related testing frequency}

We demonstrated that insufficient PrEP-related testing leads to an increase in STI prevalence as PrEP uptake increases. This led us to question whether a minimum PrEP-related testing frequency, $\lambda_{P,\text{min}}$, exists, which is necessary to decrease STI prevalence. The parameter $\lambda_P$ represents the number of mandatory tests per year for individuals taking PrEP. In the main analysis, we considered three specific values (1, 2, and 4 tests per year). Here, we adopt a more qualitative approach to explore whether a minimum required testing frequency, $\lambda_{P,\text{min}}$, emerges as a function of PrEP uptake ($P$) and risk awareness ($H$).

\begin{figure}[!hb]
    \centering
    \includegraphics[width=165mm]{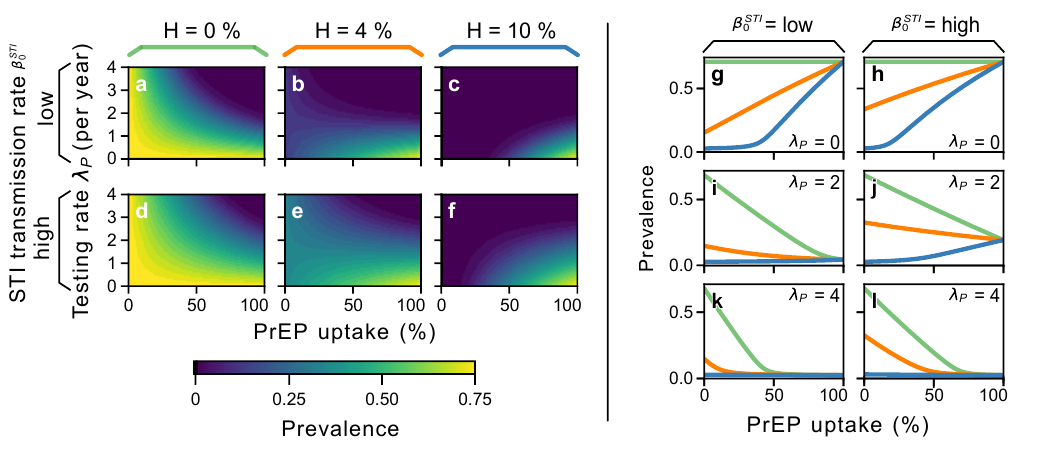}
    \caption{\textbf{Asymptomatic screening frequency among PrEP users is key to lowering STI prevalence.} STI prevalence depends on the PrEP uptake,  the PrEP-related testing rate $\lambda_P$, and risk awareness $H$ (\textbf{a-f}). At high PrEP-related testing frequencies ($\lambda_P=4$), STI prevalence never increases with increasing PrEP uptake for all levels of risk awareness (\textbf{a, d, k, l}). However, when PrEP-related testing frequencies $\lambda_P$ is below the risk-related testing frequency, this relationship becomes non-monotonic at moderate or high risk awareness (here, $H \gtrapprox 4$), and more PrEP uptake can paradoxically lead to increased STI prevalence (\textbf{b, c, e, f, g-j}). This effect arises because PrEP users undergo mandatory PrEP-related testing only; thus, when $\lambda_P$ is lower than the risk-related screening frequency, the overall screening decreases as PrEP uptake increases, resulting in higher STI prevalence. (Parameters in Tab.~\ref{tab:parameters})}
    \label{fig:min_test_frequ} 
\end{figure}

For a fixed PrEP uptake $P$ and risk awareness $H$, increasing the PrEP-related testing frequency $\lambda_P$ consistently reduces STI prevalence, as expected (Fig.~\ref{fig:min_test_frequ}a-f). For example, for high STI transmission, $P=65\,\%$, and $H= 0\,\%$, increasing $\lambda_P$ from 1/year to 4/year decreases prevalence by $88.47$\,\% (from $0.581$ to $0.067$). This suggests that more testing is always better when PrEP uptake and risk awareness are constant.   
However, when risk awareness $H$ is moderate or high (here, $H\gtrapprox4$), insufficient PrEP-related testing leads to an increase in STI prevalence as PrEP uptake rises (Fig.~\ref{fig:min_test_frequ}b,c,e,f). This occurs because individuals on PrEP undergo mandatory PrEP-related testing but not additional risk-related testing. This means that when the PrEP-related testing frequency is lower than the risk-related testing frequency, overall testing decreases as more people adopt PrEP, driving up STI prevalence. 
From these observations, we conclude that a minimum PrEP-related testing frequency, $\lambda_{P,\text{min}}$, exists and depends on risk awareness. 

In summary, insufficient PrEP-related testing frequency can contribute to a rise in STI prevalence as PrEP uptake in the population increases. Consequently, a minimum testing frequency, $\lambda_{P,\text{min}}$, exists, which depends on risk awareness. However, testing above this threshold is always beneficial and leads to a further reduction in STI prevalence. 

\subsection{Impact of asymptomatic fraction on disease dynamics}
\label{SI_sec:assymptomatic_rate_0.7}

In the main text, we used an asymptomatic fraction of $\psi=0.85$ \cite{lamontagne2003chlamydia,korenromp2002proportion} (see Tab. \ref{tab:parameters}). To assess the impact of this parameter, we performed a sensitivity analysis in which the asymptomatic fraction was reduced to $0.7$, following the value suggested by William \textit{et al.}\cite{williams2023frequent}.

\begin{figure}[!hb]
    \centering
    \includegraphics[width=165mm]{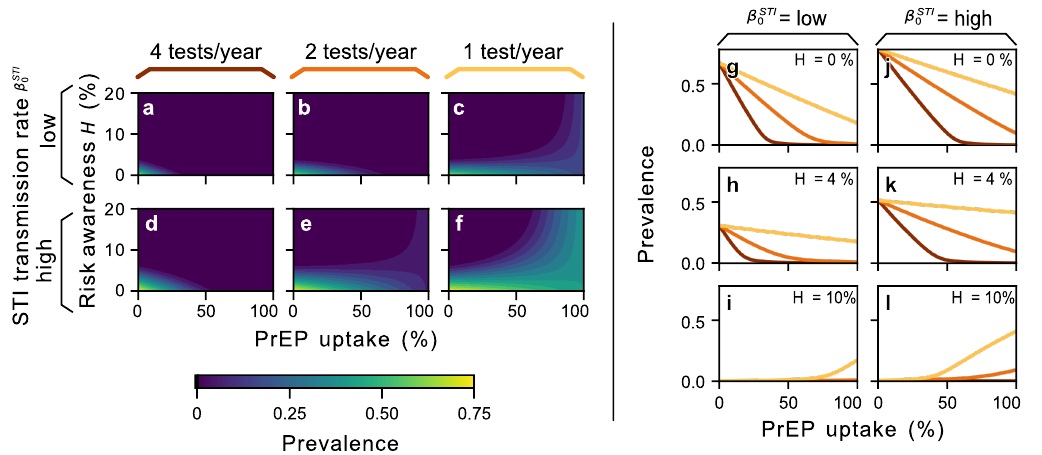}
    \caption{\textbf{Increasing PrEP uptake can mitigate the spread of chlamydia.} We analyze the prevalence of chlamydia infections under two scenarios of transmission rates (low and high; parameters in Tab.~\ref{tab:parameters}) for an asymptomatic fraction of $\psi=0.7$ (in the main text we used $\psi=0.85$). The frequency of PrEP-related testing determines the effects of PrEP uptake, particularly at low levels of risk awareness in the population ($H$).}
    \label{fig:figure_1_psi_0.7} 
\end{figure}

Our simulations indicate that this reduction in the asymptomatic fraction does not lead to any qualitative changes in the overall epidemic dynamics; the trends observed in the main analysis remain consistent (Fig. \ref{fig:figure_1_psi_0.7}). However, as expected, a lower asymptomatic fraction decreases overall prevalence compared to the baseline scenario (cf. Fig. \ref{fig:results_1}). This outcome is not surprising; individuals are assumed to seek testing and treatment voluntarily as they develop symptoms, removing themselves earlier from the pool of infection should that be the case. 

\clearpage
\subsection{Testing artifacts with rising risk awareness}
\label{SI_sec:Nobs_Nreal_riskawareness}

\begin{figure}
    \centering
    \includegraphics[width=0.5\linewidth]{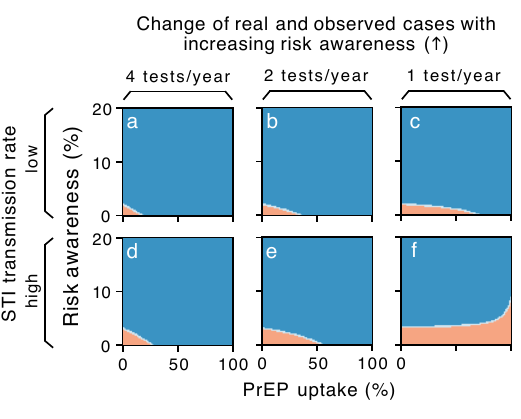}
    \caption{\textbf{Comparison of the dynamics of observed and real cases.} An increase in PrEP uptake will change the prevalence \Nreal and observed cases \Nobs. Both can either increase or decrease, resulting in four different combinations. These can be categorized into 3 categories: (light blue) both \Nreal and \Nobs will increase, (dark blue) both \Nreal and \Nobs will decrease, (red) one of the two increases, while the other decreases (in our case, we only observe the case where \Nreal decreases while \Nobs increases). While the positive tests mirror the real dynamics most of the time, we found some regions with mismatches (red) where real cases decrease while observed cases rise. (Parameters in Tab.~\ref{tab:parameters})}
    \label{SI_fig:smallmodel_Nobs_Nreal_riskawareness}
\end{figure}

In the main text, we looked at the changing dynamics of STI prevalence \Nreal and positive tests \Nobs when PrEP uptake changes. Here, we will look at changing risk awareness in the population (Fig.~\ref{SI_fig:smallmodel_Nobs_Nreal_riskawareness}).

The risk awareness represents how people perceive their risk of infection on average and undergo voluntary testing even without symptoms. Increasing risk awareness in the population, therefore, leads to increased testing and in our case lowers \Nreal. If both risk awareness and PrEP uptake are low, this increased testing decreases \Nreal at such a rate that \Nobs can increase; although the fraction of positive tests declines, the overall volume of testing rises enough that the absolute number of positive results still increases. If PrEP uptake stays low and risk awareness increases further, however, both \Nobs and \Nreal decrease. The real cases now decrease so much that even though we test more, fewer positive tests are recorded. When PrEP uptake is high, most people already undergo PrEP-related testing, and the increase in overall testing due to increased risk awareness is only small, so \Nobs also decreases here. Only for a low PrEP-related testing rate and high transmission rate (Fig.~\ref{SI_fig:smallmodel_Nobs_Nreal_riskawareness}f), \Nreal decreases so slowly that increased testing leads to increased \Nobs; the paradoxical region has a larger extent, supporting the previous result that frequent testing for PrEP users is essential for effective containment. Overall, increasing risk awareness in the population has not only the positive effect of decreased \Nreal but also leads to \Nobs correctly reflecting the decreasing trend of \Nreal.

\clearpage
\subsection{Extended model with risk group stratification}\label{SI_sec:extended_model}

While our main analysis is based on a homogeneous population model, we also considered an extended model with risk group stratification to better capture real-world heterogeneity. This stratified model includes four distinct risk groups, with their relative sizes and interaction patterns informed by an HIV model for high-risk MSM using data from the Netherlands \cite{rozhnova2018elimination}. The four risk groups comprise $45.1\%$, $35.3\%$, $12.5\%$, and $7.1\%$ of the total MSM population, respectively \cite{rozhnova2016impact}. We denote the groups by indices $l \in \{1, 2, 3, 4\}$, ordered by increasing risk: group $l = 1$ represents individuals at the lowest risk, while group $l = 4$ includes those at the highest risk. Parameters that differ from the small model (Tab. \ref{tab:parameters}) are listed in Table \ref{tab:sensitivityanalysis}.

\begin{align*}
\frac{dS_l}{dt} &= 
    - \Lambda_l\, S_l
    + \gamma\, I^{a}_l
    + \tilde{\gamma} T_l
    + \Sigma +\Phi
    - \mu\, S_l, \\
\frac{dI^{a}_l}{dt} &=
    \psi\, \Lambda_l\, S_l
    - \gamma\, I^{a}_l
    - \lambda_{a,l} (P,H)\, I^{a}_l
    - \mu\, I^{a}_l
    + \psi\, \Sigma, \\
\frac{dI^{s}_l}{dt} &=
    (1-\psi)\, \Lambda_l\, S_l
    - \lambda_{s,l} (P,H)\, I^{s}_l
    - \mu\, I^{s}_l
    + (1-\psi)\, \Sigma,\\
\frac{dT_l}{dt} &=
    \lambda_{a,l} (P,H)\, I^{a}_l
    + \lambda_{s,l} (P,H)\, I^{s}_l
    - \tilde{\gamma} T_l
    - \mu\, T_l,
\end{align*}

\bigskip

\noindent
where
\begin{align*}
&\Lambda_l = \beta^{STI}_{0} \left( 1 - m(H)\,(1-P) \right)
    \sum_{l'} M_{ll'} \frac{I^{a}_{l'} + I^{s}_{l'}}{N_{l'}}, \\
&m(H) = m_{\min} + (m_{\max} - m_{\min})
    \left( 1 - \exp\left(-\frac{H}{H_{\mathrm{thres}}} \right) \right), \\
&M_{ll'} = \omega\, \frac{c^{\mathrm{HIV}}_{l'}\, N_{l'}}{\sum_k c^{\mathrm{HIV}}_k\, N_{k}} + (1-\omega) \delta_{ll'}, \qquad\\
&\lambda_{a,l} (P,H) = k_{\text{or}, l}\, \beta^{HIV}_0\, \left(1-m(H)\right)\, H\, (1-P) + \lambda_P\, P, \\
&\lambda_{s,l} (P,H) = \lambda_0 + \lambda_a (P,H)\, .
\end{align*}

$N_l$ is the population size in group $l$, and $\delta_{ll'}=1$ if $l=l'$, otherwise $\delta_{ll'}=0$.

The increased dimensionality and complexity introduced by risk group stratification made analytical calculation of the model’s fixed points highly challenging. As a result, directly solving for equilibrium states was impractical.

To address this, we employed a numerical approach, simulating the model dynamics over a sufficient number of years to allow the system to evolve towards a stable equilibrium. Equilibrium was operationally defined as the point at which the values of all relevant state variables--namely, the fraction of susceptible ($S_l$), asymptomatically infected ($I^{a}_l$), symptomatically infected ($I^{s}_l$), and treated ($T_l$)--changed by less than a small threshold $\epsilon$ between consecutive time steps, i.e., $|x^{(t)} - x^{(t-1)}| \leq \epsilon \,\, \text{for all} \,\, x \in \{S_l, I^{a}_l, I^{s}_l, T_l\}$. This stopping criterion ensured that the model output reflected steady-state conditions, minimizing the influence of initial conditions or transient dynamics. All results from the risk-stratified extended model, as presented in this study, were obtained from simulations in equilibrium.

\subsection{Sensitivity Analysis of the extended model with risk group stratification}

To assess the robustness of our model, we performed a sensitivity analysis focusing on the odds ratios ($k_\text{or}$) assigned to the different risk groups ($l$). Specifically, we examined three scenarios with distinct combinations of odds ratios for the four risk groups \cite{kesler2016actual} (see Tab. \ref{tab:sensitivityanalysis}). To ensure comparability between scenarios, all sets of odds ratios were normalized so that the population-weighted mean odds ratio remained fixed at $k_\text{or}=50$. This normalization isolates the effect of risk distribution while holding the average risk in the population constant.

\begin{figure}
    \centering
    \includegraphics[width=\linewidth]{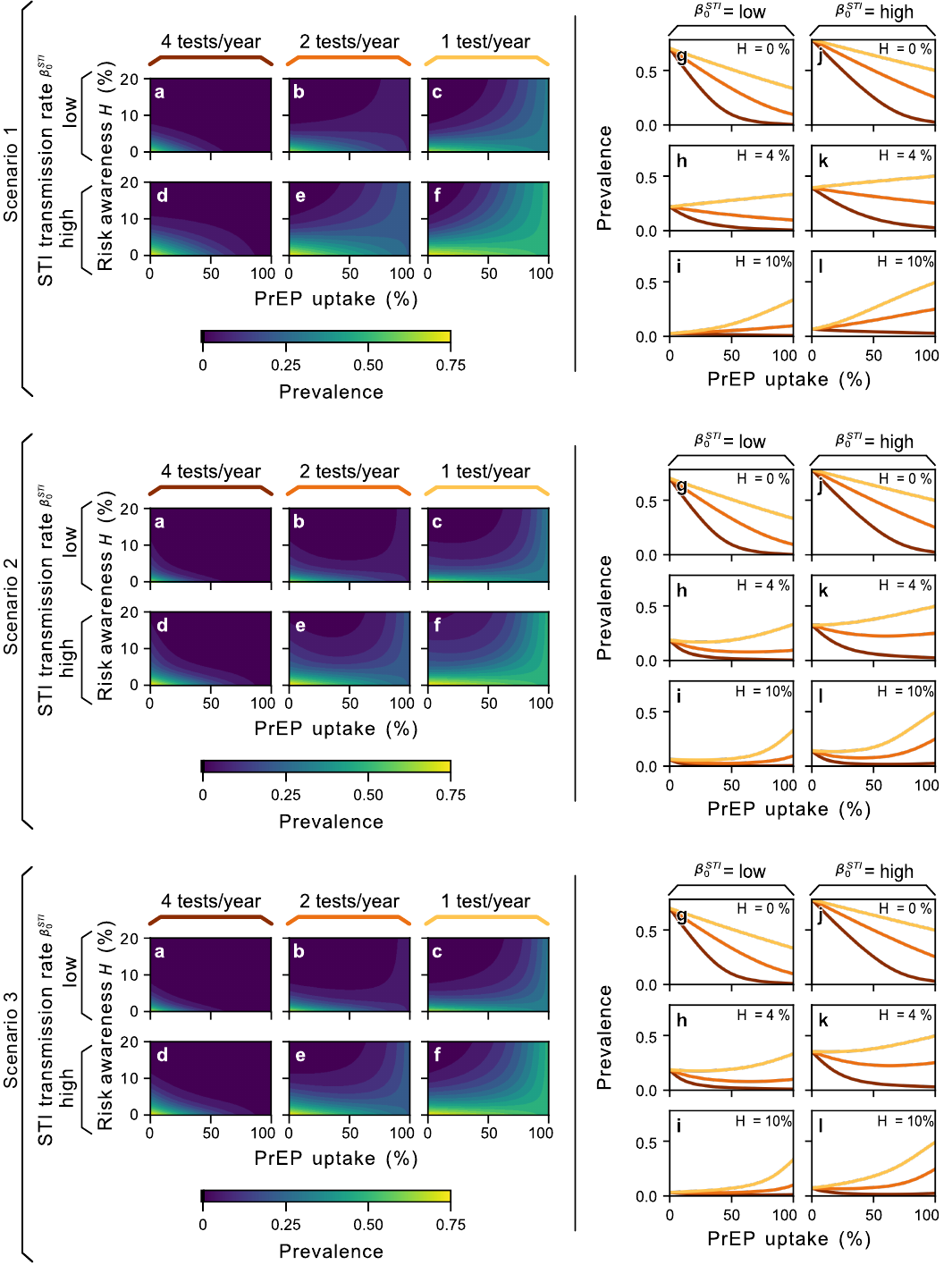}
    \caption{Sensitivity analysis with varying odds ratios among the risk groups in the extended model (for the exact value for each scenario see Tab.~\ref{tab:sensitivityanalysis}) (mirroring Fig.~\ref{fig:results_1})}
    \label{SI_fig: figure1_scenarios123}
\end{figure}

\begin{figure}
    \centering
    \includegraphics[width=\linewidth]{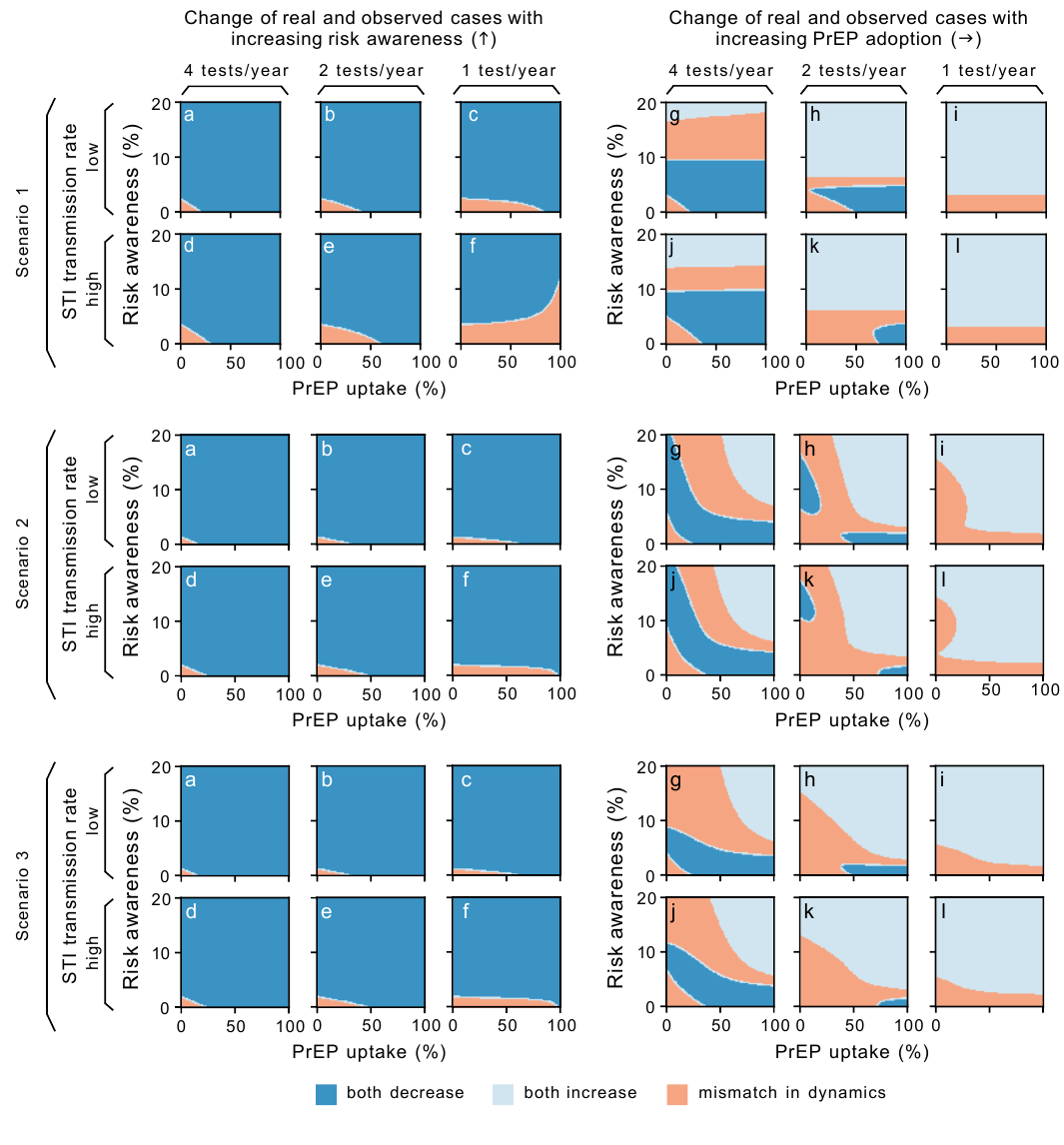}
    \caption{Sensitivity analysis with varying odds ratios among the risk groups in the extended model (for the exact value for each scenario see Tab.~\ref{tab:sensitivityanalysis}) (mirroring Fig.~\ref{fig:results_2})}
    \label{SI_fig: figure2_scenarios123}
\end{figure}

For each scenario, we reproduced the main results (Fig. \ref{fig:results_1} and \ref{fig:results_2}) using the corresponding odds ratio distributions. While we observed that the absolute values of prevalence differed notably across the three scenarios, the overall qualitative dynamics and trends remained similar to those presented in the main figures (Fig. \ref{SI_fig: figure1_scenarios123} and \ref{SI_fig: figure2_scenarios123}). This indicates that, although the magnitude of model outputs may vary with different risk heterogeneity assumptions, the general patterns and qualitative conclusions of our analysis are robust to changes in the configuration of risk group odds ratios.

\begin{table*}[ht]\caption{Model parameters of the extended model with risk group stratification}\label{tab:sensitivityanalysis}
\centering
\begin{tabular}{lp{6.5cm}p{4.5cm}ll}
\toprule
Parameter & Definition   & Value & Units & Source \\
\midrule
$\omega$ & Mixing parameter & $0.5$ & -- & \cite{rozhnova2018elimination}\\
$c^{\mathrm{HIV}}_{l}$ & Average number of new partners per year in risk group $l$ & $[0.13, 1.43, 5.44, 18.21]$ & \SI{}{yr^{-1}} & \cite{rozhnova2018elimination}\\
$k_{or, l}$ & Odds ratio in risk group $l$, likelihood that one perceives risk if one is at risk &
\begin{tabular}[t]{@{}l@{}}
Scenario 1: $[50,  50,   50,  50]$ \\
Scenario 2: $[2,  63.5,  100,  200]$ \\
Scenario 3: $[14,  24,  167,  203]$
\end{tabular}
& -- & \cite{kesler2016actual} \\
\bottomrule
\end{tabular}
\end{table*}

\clearpage

\end{document}